\documentclass[prl,aps,twocolumn,superscriptaddress,bibnotes,footinbib]{revtex4-2}
\usepackage{amsmath,amssymb}
\usepackage[colorlinks,bookmarks=false,citecolor=magenta,linkcolor=magenta,urlcolor=magenta]{hyperref}
\usepackage{tikz}
\usepackage{mathtools}
\usepackage{bm}
\usepackage{ragged2e}
\usepackage{graphicx}
\usepackage[symbol]{footmisc}
\newcommand{\mC}{\mathcal C}
\newcommand{\mL}{\mathcal L}

\newcommand{\myroman}[1]{\uppercase\expandafter{\romannumeral#1}}
\makeatletter
\newsavebox{\@brx}
\newcommand{\llangle}[1][]{\savebox{\@brx}{\(\m@th{#1\langle}\)}%
  \mathopen{\copy\@brx\kern-0.5\wd\@brx\usebox{\@brx}}}
\newcommand{\rrangle}[1][]{\savebox{\@brx}{\(\m@th{#1\rangle}\)}%
  \mathclose{\copy\@brx\kern-0.5\wd\@brx\usebox{\@brx}}}
\makeatother
\newcommand{\figpanels}[3]{Fig.~\hyperref[#1]{\ref*{#1}(#2)-(#3)}}

\begin{document}
	\title{Quantum synchronization in one-dimensional topological systems}
	\author{Tong Liu}
	\email{tliu.phys@gmail.com}
	\affiliation{Department of Microtechnology and Nanoscience, Chalmers University of Technology, 41296 Gothenburg, Sweden}

	\author{Laura García-Álvarez}
	\affiliation{Department of Microtechnology and Nanoscience, Chalmers University of Technology, 41296 Gothenburg, Sweden}

	\author{Giovanna Tancredi}
	\affiliation{Department of Microtechnology and Nanoscience, Chalmers University of Technology, 41296 Gothenburg, Sweden}

	\begin{abstract}

			The phenomenon of synchronization, where entities exhibit stable oscillations with aligned frequencies and phases, has been revealed in diverse areas of natural science. 
			It plays a crucial role in achieving frequency locking in multiple applications such as microwave communication and signal processing.
			The study of synchronization in quantum systems has gained significant interest, particularly in developing robust methods for synchronizing distant objects.
			Here, we demonstrate that synchronization between the boundary sites of one-dimensional generalized Aubry-Andr\'e-Harper models can be induced through applying dissipation on the central sites.
			We observe two types of synchronization, stemming from the topological edge states, identified by the off-diagonal or diagonal correlations between the boundary sites.
			We calculate the relaxation rate to realize the synchronization and its acceleration with bulk dissipation.
			Remarkably, the synchronous oscillations maintain steady amplitude and frequency in the thermodynamic limit.
			Moreover, we show that the synchronization is robust against perturbations in the Hamiltonian and initial states, highlighting its potential for practical implementation on near-term quantum simulation platforms.

	\end{abstract}

	\maketitle

	Synchronization is a universal classical dynamical phenomenon observed across various fields such as physics, biology, and engineering~\cite{Pikovsky2001,Boccaletti2018}.
	It typically manifests in nonlinear systems when individual frequencies or phases become locked owing to an external periodic drive, mutual coupling between subsystems or stochastic noise~\cite{vanderPol, RevModPhys.77.137,PhysRevLett.88.230602, PhysRevLett.93.204103}.
	This phenomenon has found broad applications in wireless communication~\cite{Bregni2002}, signal processing~\cite{Strogatz}, and neuro-inspired computing~\cite{Hoppensteadt2001}.

	Recently, the study of synchronization has been extended into the quantum realm with numerous advances in both theoretical frameworks and experiment demonstrations 
	~\cite{PhysRevLett.107.043603,PhysRevLett.109.233906,PhysRevLett.111.103605,PhysRevLett.111.213902,PhysRevLett.112.094102,PhysRevLett.114.113602,PhysRevLett.115.163902,Weiss2016,PhysRevLett.118.063605,PhysRevLett.120.163601,
	PhysRevLett.121.053601,PhysRevLett.121.063601,PhysRevLett.125.013601,
	PhysRevLett.111.234101,PhysRevA.91.061401,PhysRevResearch.5.023021,PhysRevResearch.5.033209,
	PhysRevA.97.013811,PhysRevResearch.2.023026,PhysRevLett.132.196601,PhysRevLett.133.020401}.
	Compared to their classical counterparts, quantum systems exhibit more complex synchronization behaviors.
	In quantum van der Pol (vdP) oscillators, quantized energy levels can enhance phase locking in the presence of strong nonlinear damping~\cite{PhysRevLett.111.234101}.
	Conversely, quantum noise may reduce the frequency entrainment of a quantum vdP oscillator subject to a weak driving~\cite{PhysRevLett.112.094102}.
	Moreover, the introduction of a large Kerr anharmonicity leads to phase synchronization at multiple resonant frequencies, a phenomenon absent in classical systems~\cite{PhysRevLett.117.073601}.
	Additionally, parametric (two-photon) driving can achieve stronger synchronization than coherent driving~\cite{PhysRevLett.120.163601}.
	As a unique tool in quantum systems, measurement can also induce synchronization in a continuously monitored system~\cite{PhysRevLett.132.010402}.
	However, most efforts have been so far focused on synchronization within systems composed of a few oscillators or spins.  
	Observing quantum synchronization in many-body systems is challenging due to several obstacles.
	For instance, the oscillation amplitudes of observables employed to characterize synchronization may vanish in the thermodynamic limit~\cite{PhysRevLett.113.154101,PhysRevLett.129.250601,PhysRevLett.132.010402}, which restricts their applicability in macroscale networks.
	Furthermore, while collective synchronization can arise in ensembles of globally coupled systems~\cite{PhysRevLett.107.043603,PhysRevLett.111.234101,PhysRevLett.115.163902,PhysRevLett.131.190402}, scaling such systems in experiments presents significant difficulties~\cite{Liu2023,Zhang2023}.

	Here, we address these challenges by demonstrating noise-induced synchronization in the Aubry-Andr\'e-Harper (AAH) model and its generalizations, widely studied in the contexts of localization and topological states~\cite{PhysRevLett.50.1873,PhysRevLett.51.1198,PhysRevLett.108.220401,PhysRevLett.109.106402,PhysRevLett.109.116404}.
	We characterize two types of synchronization between remote edge sites in a long chain, consisting of over 100 sites, by applying noise to the central sites.
	We reveal that the chiral and reflection symmetries guarantee that the populations at the far ends synchronously oscillate.
	We show that the amplitudes and frequencies of the population oscillation at the boundary sites are stable in the thermodynamic limit, even in the absence of global interactions.
	We calculate the lowest relaxation rate for synchronization and examine how bulk dissipation can accelerate the relaxation without disrupting synchronization.
	We finally illustrate that the synchronization is robust under perturbations of both the Hamiltonian and initial states, which is built upon the topological nature of edge states.

	\emph{Synchronization of off-diagonal correlations.---}
	We begin by illustrating the first type of synchronization in the generalized 1D AAH model, which arises without both chiral and reflection symmetries.
	The model is described by the Hamiltonian
	\begin{equation}
		H = \sum_{j=1}^N V_jn_j + \sum_{j=1}^{N-1}\left(g_j c_{j+1}^\dagger c_j + \mathrm{h.c.}\right)\label{eq:generalized_AAH},
	\end{equation}
	where $N$ is the number of sites, $c_j$ ($c_j^\dagger$) is the fermionic annihilation (creation) operator at site $j$, $n_j$ is the number operator at site $j$, $g_j = g[1 + \lambda \cos(2\pi\alpha j + \phi_\lambda)]$ is the hopping strength between site $j$ and site $(j+1)$, and $V_j=V\cos(2\pi\alpha j + \phi_V)$ is the on-site potential energy at site $j$.
	Both the hopping strength and the on-site potential energy are modulated by cosine functions with the same period $1/\alpha$ and respective phases $\phi_\lambda$ and $\phi_V$.
	In the following context, $\alpha$ is always rational and can be expressed as $\alpha=p/q$ with $p$ and $q$ being co-prime integers.
	The special case $\lambda=0$ reduces to the diagonal AAH model which could be derived from the Hamiltonian in the $x$ direction of a 2D quantum Hall (QH) model by imposing a periodic boundary condition in the $y$ direction~\cite{PhysRevLett.45.494,PhysRevLett.49.405}.
	The good quantum number, momentum in the $y$ direction, degenerates into the diagonal phase $\phi_V$, which assumes values from the first Brillouin zone (1BZ).
	Since the on-site potential is periodic with a period $q$, the bulk wavefunction takes the Bloch form and bulk energies decompose into $q$ bands.
	We start with the case of $p=1$ and $q=3$ leading to two edge states, which facilitates the long-range synchronization between edge sites.

\begin{figure}[h]
		\centering
		\includegraphics[width=0.96\linewidth]{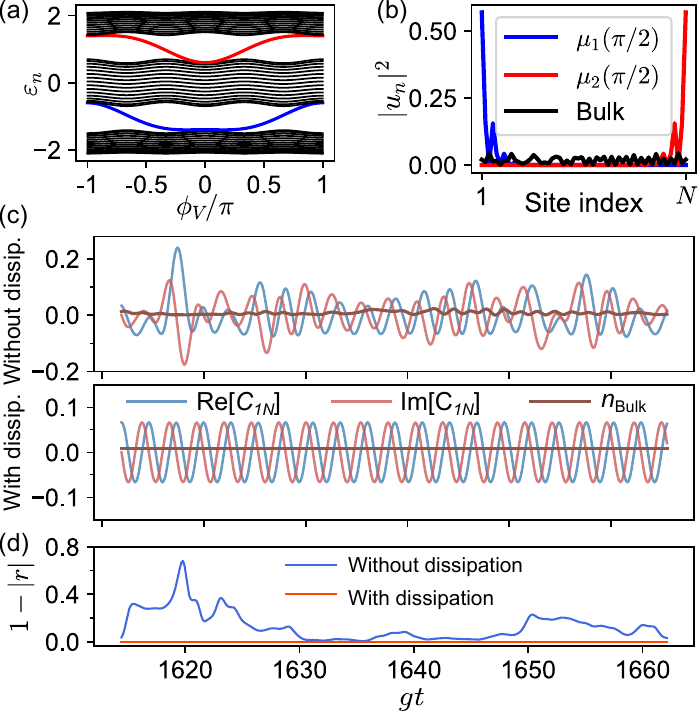}
		\caption{\quad Off-diagonal synchronization in the diagonal AAH model. 
		(a) Energy spectrum of bulk energies (black) and edge states (red, blue).
		(b) Amplitudes of two edges states at $\phi_V=\pi/2$ and a bulk state in the middle of the band. 
		(c) Evolution of the two-site correlation function $\mathcal C_{1N} = \langle c_1^\dagger c_N \rangle$ and central site density without dissipation (upper panel) and with $\gamma/g=1.5$ (lower panel) where $N=59$, $v=0.7$, and $\phi_V=\pi/2$.
		(d) Pearson coefficients between $\mathrm{Re}[\mathcal C_{1N}]$ and $\mathrm{Im}[\mathcal C_{1N}]$ after a phase shift.
		}\label{fig:off-diagonal}
	\end{figure}

	Suppose that the $n$th eigenstate of the single-particle Hamiltonian of Eq.~(\ref{eq:generalized_AAH}) is given by $|\psi_n\rangle=\sum_ju_{j,n}c_j^\dagger|0\rangle$ and $N=ql-1$ where $l$ is a positive integer, the eigenvalue equation leads to the following Harper equation
	\begin{equation}
		gu_{j+1,n} + gu_{j-1,n} + V\cos(2\pi\alpha j + \phi_V)u_{j,n} = E_n u_n,\label{eq:Harper}
	\end{equation}
	where $u_{j,n}$ is the amplitude of the wavefunction at site $j$ and $E_n$ is the $n$th single particle energy.
	As illustrated in Fig.~\ref{fig:off-diagonal}(a)-(b), two edge states are located within the top and bottom gaps.
	The edge energies $\mu_1$ and $\mu_2$ are given by 
	\begin{align}
	\mu_1(\phi_V)/g = -v\cos(\phi_V)/2 - \sqrt{1+3v^2\sin^2(\phi_V)/4}, \notag \\
	\mu_2(\phi_V)/g = -v\cos(\phi_V)/2 + \sqrt{1+3v^2\sin^2(\phi_V)/4},
	\end{align}
	with $v = V/g$.
	After straightforward calculations, we find that the edge state corresponding to $\mu_1$ $(\mu_2)$ is localized at the right (left) edge when $\phi_V\in (-\pi,0)$ and at the left (right) edge when $\phi_V\in (0,\pi)$~\cite{sm}.
	Therefore, the two edge states always reside at opposite edges for any value of $\phi_V$.

	To achieve synchronization between edge states, we introduce local dissipation targeted at sites $\mathcal S$.
	The density matrix of the system $\rho$ follows the Lindblad master equation $\dot\rho(t) = \mathcal L(\rho) = -i[H,\rho] + \gamma\sum_{s\in\mathcal S}(J_s\rho J_s^\dagger - 1/2\{J_s^\dagger J_s, \rho\})$ where $\gamma$ is the dissipation strength, $J_s$ is the jump operator at site $s$, and $\mL$ is the corresponding Lindblad superoperator~\cite{Lindblad1976,Breuer2007}.
	For simplicity, we choose $J_s$ as the number operator.
	Explicitly quantifying the synchronization involves considering the two-site correlation function $\mC_{ij}(t)\equiv \langle c_i^\dagger c_j(t)\rangle$ where the diagonal terms describe the average on-site population.
	A straightforward calculation shows that the dynamics of $\mC$ also follows a Lindblad master equation, which is employed to numerically compute the evolution of $\langle c_i^\dagger c_j\rangle$ investigated in our work~\cite{sm}.
	Using the spectral decomposition of $\mL$, the evolution of $\mC$ is given by $\mathcal C(t) = \sum_k e^{\lambda_k t} |R_k\rrangle\llangle L_k|\mathcal C(0)\rrangle$, 
	 where $\lambda_k$ is the eigenvalue of $\mL$, $|R_k\rrangle$ ($|L_k\rrangle$) is the right (left) eigenoperator of $\mL$, and the inner product $\llangle A| B\rrangle$ between two operators $A$ and $B$  is defined as $\mathrm{Tr}(A^\dagger B)$~\cite{Buca2022}.
	 Stable synchronization occurs when all the real parts of the eigenvalues are negative, except for a conjugate imaginary pair $\lambda_1 = i(\varepsilon_m - \varepsilon_n)$ and $\lambda_2=\lambda_1^*$ where $\varepsilon_m$ and $\varepsilon_n$ are eigenenergies of the Hamiltonian $H$~\cite{PhysRevLett.129.250601}.
	 After the other modes decay to zero, the system is confined to the subspace spanned by $\{|\psi_m\rangle\langle\psi_m|,|\psi_m\rangle\langle\psi_n|,|\psi_n\rangle\langle\psi_m|, |\psi_n\rangle\langle\psi_n|\}$ where $|\psi_m \rangle$ and $|\psi_n \rangle$ are the eigenstates corresponding to $\varepsilon_m$ and $\varepsilon_n$, respectively.
	 The evolution of $\mC_{ij}(t)$ in the subspace is described by
	 \begin{equation}
	 	\mC_{ij}(t) = u_{i,m}u_{j,n}c_0e^{i\omega_{mn}t} + u_{i,n}u_{j,m}c_0^*e^{-i\omega_{mn}t},\label{eq:correlation}
	 \end{equation}
	 up to a constant where $c_0=\langle\psi_m|\mC(0)|\psi_n\rangle$ and $\omega_{mn}\equiv|\varepsilon_m-\varepsilon_n|$.

	 By specifying the noise as on-site dephasing at the two centermost sites, i.e., $\mathcal S = \{N/2,N/2+1\}$, only two edge modes are immune to the dissipation, thereby constituting a decoherence-free subspace when $N\rightarrow\infty$.
	 Figure~\ref{fig:off-diagonal}(c) shows the evolution of off-diagonal correlations between boundary sites $\mathrm{Re}[\mathcal C_{1N}] = \langle( c_1^\dagger c_N + c_N^\dagger c_1)/2\rangle$ and $\mathrm{Im}[\mathcal C_{1N}]=\langle(c_1^\dagger c_N - c_N^\dagger c_1)/2i\rangle$. 
	 Analogous to the diagonal correlations $\langle c_1^\dagger c_1\rangle$ and $\langle c_N^\dagger c_N\rangle$, it is natural to explore whether synchronization exists between these off-diagonal functions.
	 The initial state is chosen as a product state $|+00 \cdots 0+\rangle$ with $|+\rangle_j=(|0\rangle_j+|1\rangle_j)/\sqrt 2$ where $|0\rangle_j$ and $|1\rangle_j$ denotes the vacuum state and excitation state at site $j$, respectively.
	 As a comparison, the upper panel depicts the free evolution of $\mathrm{Re}[\mathcal C_{1N}]$ and $\mathrm{Im}[\mathcal C_{1N}]$ in the absence of dissipation where the oscillations are out of phase and exhibit the superposition of different modes.
	 In contrast, the lower panel illustrates that after dissipation is applied, $\mathrm{Re}[\mathcal C_{1N}]$ and $\mathrm{Im}[\mathcal C_{1N}]$ synchronize with a constant phase difference of $\pi/2$. 
	 This synchronization occurs because $\mathcal C_{1N}(t)$ becomes proportional to $e^{i2|\mu_1(\pi/2)-\mu_2(\pi/2)|t}$ up to a constant, as described by Eq.~(\ref{eq:correlation}).
	 The synchronization can be confirmed by the Pearson coefficient which is defined as $r[f,h](t)=\mathrm{Cov}[f,h]/\sqrt{\mathrm{Var}[f]\mathrm{Var}[h]}$ for two time-dependent functions $f(t)$ and $h(t)$~\cite{PhysRevA.85.052101,PhysRevA.88.042115,Tindall2020,PhysRevLett.129.250601}.
	 Synchronized oscillations lead to $|r|=1$ while the uncorrelated functions imply $r=0$.
	 Figure~\ref{fig:off-diagonal}(d) plots the Pearson coefficient $r$ between $\mathrm{Re}[\mathcal C_{1N}(t)]$ and $\mathrm{Im}[\mathcal C_{1N}(t+\tau)]$ where $\tau = \pi/2\omega$ is the time shift calculated by the theoretical frequency to align the phases.
	 The Pearson coefficient converging to one in the case with dissipation further confirms that the oscillation frequency matches the theoretical result. 

	\emph{Synchronization of diagonal correlations.---}
	We have demonstrated that two QH edge states enable off-diagonal correlations between edge sites.
	In practice, it is preferable to observe synchronization in diagonal correlations or local on-site populations.
	In the following, we show that such synchronization can be observed in the AAH model by incorporating chiral symmetry and reflection symmetry~\cite{sm}.
	We now consider the off-diagonal AAH model corresponding to $V=0$ and $\lambda\ne 0$ in Eq.~(\ref{eq:generalized_AAH}).
	When $\alpha$ takes the value of $1/2$, Majorana modes emerge on this model, which is similar to the Kitaev chain, attributed to the additional chiral symmetry~\cite{PhysRevLett.110.180403}.
	Here we focus on the case $\alpha = 1/4$, i.e., $p=1$ and $q=4$, where the chiral symmetry is also preserved.

	\begin{figure}[t]
		\centering
		\includegraphics[width=0.96\linewidth]{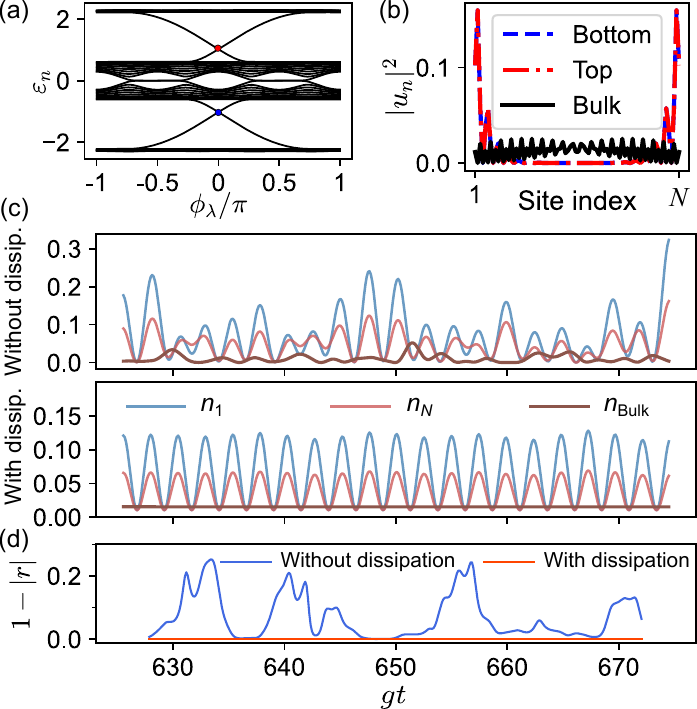}
		\caption{\quad Diagonal synchronization in the off-diagonal AAH model. 
		(a) Energy spectrum. Red and blue circles indicate the degenerate points of the edge states. 
		(b) Amplitudes of edges states at degenerate points and a bulk state in the middle of the band.
		(c) Density evolution at edge and middle sites without dissipation (upper panel) and with $\gamma/g=2$ (lower panel) where $N=80$, $\lambda=0.2$, and $\phi_\lambda=0$.
		(d) Pearson coefficients between $n_1$ and $n_N$.}\label{fig:diagonal}
	\end{figure}

	\begin{figure}[t]
		\centering
		\includegraphics[width=0.96\linewidth]{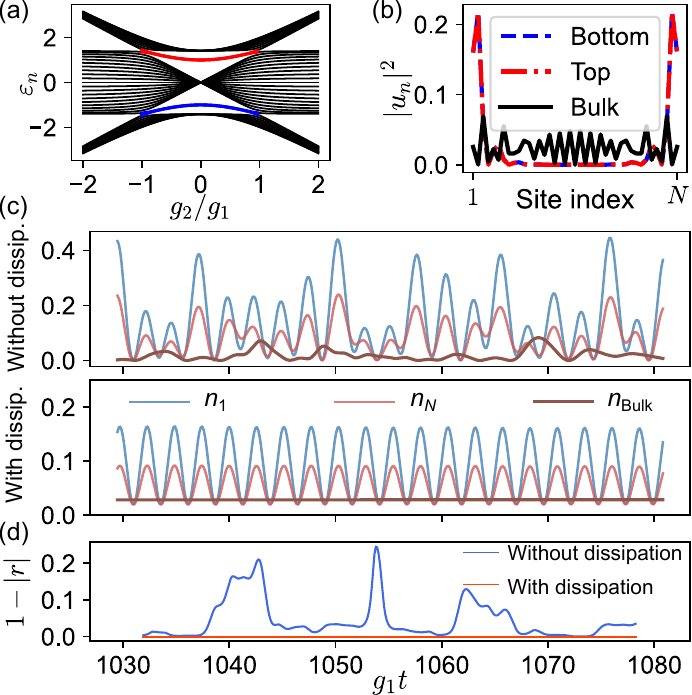}
		\caption{\quad Diagonal synchronization in the generalized four-band off-diagonal AAH model. 
		(a) Energy spectrum of bulk states (black) and degenerate edge states (red, blue). 
		(b) Amplitudes of the edges states and a bulk state in the middle of the band with $g_2/g_1=0.7$.
		(c) Density evolution at edge and middle sites without dissipation (upper panel) and with $\gamma/g_1 = 2$ (lower panel) where $N=41$ and $g_2/g_1=0.7$.
		(d) Pearson coefficients between $n_1$ and $n_N$.}\label{fig:generalized}
	\end{figure}

	Figure~\ref{fig:diagonal}(a) shows the normalized energy for $\phi_\lambda$ taking the value from 1BZ where $N=4l$ with an open boundary condition and $\mathcal S = \{N/2, N/2+1\}$.
	The top and bottom bands in the four bands are fully gapped which indicates the existence of QH edge states. 
	However, the central two bands are gapless and two zero-energy edge modes are found for $-3\pi/4 < \phi_\lambda < -\pi/4$ and $\pi/4<\phi_\lambda<3\pi/4$.
	In the bottom and top band gaps,
	a pair of left QH edge states with energies $\pm \sqrt{2+\lambda^2-2\sqrt 2\lambda\sin(\phi_\lambda+\pi/4)}$ emerge for $-3\pi/4 < \phi_\lambda < \pi/4$ and a pair of right QH edge states with energies $\pm \sqrt{2+\lambda^2+2\sqrt 2\lambda\sin(\phi_\lambda-\pi/4)}$ emerge for $-\pi/4 < \phi_\lambda < 3\pi/4$~\cite{sm}.
	To observe the diagonal synchronization, we require the Hamiltonian to hold a reflection symmetry [$c_j\rightarrow c_{N+1-j}^\dagger$ and $c_j^\dagger\rightarrow c_{N+1-j}$].
	It implies that $\sin(\phi_\lambda)=0$ or $\phi_\lambda=0$ ($\phi_\lambda=\pi$ is ruled out due to the absence of edge states), where four QH edge states are degenerate at energies $\pm \varepsilon^* = \pm \sqrt {2+\lambda^2-2\lambda}$.
	In the upper panel of Fig.~\ref{fig:diagonal}(c), we show the evolution of density operators located at the edge sites and the middle site without dissipation.
	The initial state is prepared as $|100\dots 0+\rangle$.
	The interference of propagation of two excitations results in the unsynchronized population fluctuation between edge sites.
	The nonzero density at the middle site also indicates the propagation of excitations over time.
	On the contrary, the populations at edge states are synchronized with the frequency $\omega=2\varepsilon^*$ under the dissipation applied to the bulk states.
	Although the oscillation amplitude at the right edge site is half of that at the left edge site due to the initial condition,
	the Pearson coefficient $r[n_1, n_N]$ shown in Fig.~\ref{fig:diagonal}(d) signifies the stable synchronization driven by the dissipation.

	To achieve synchronization over an extensive parameter region, we consider a generalized four-band off-diagonal AAH model characterized by periodic coefficients $(g_1, g_2, -g_2, -g_1)$.
	For $N = 4l+1$, the Hamiltonian holds another reflection symmetry [$c_j\rightarrow (-1)^j c_{N+1-j}^\dagger$ and $c_j^\dagger\rightarrow (-1)^j c_{N+1-j}$].
	Combined with chiral symmetry, four degenerate QH edge states emerge for $-1 < g_2/g_1 < 1$ with energies $\pm\sqrt{g_1^2+g_2^2}$ as shown in Fig.~\ref{fig:generalized}(a)~\cite{sm}.
	These four edge states collectively form the synchronization mode.
	In Fig.~\ref{fig:generalized}(b), we depict the amplitudes of edge states within the bottom gap or top gap, which resemble the edge states at $\phi_V=0$ shown in the off-diagonal AAH model.
	By specifying $\mathcal S = \{(N-3)/2, (N-1)/2, (N+1)/2, (N+3)/2, (N+5)/2\}$ and initializing the state as $|100\dots 0+\rangle$, synchronization between two edges occurs under dissipation, featuring a mutual oscillation frequency as illustrated in Fig.~\ref{fig:generalized}(c) and (d).

\begin{figure}[ht]
		\centering
		\includegraphics[width=\linewidth]{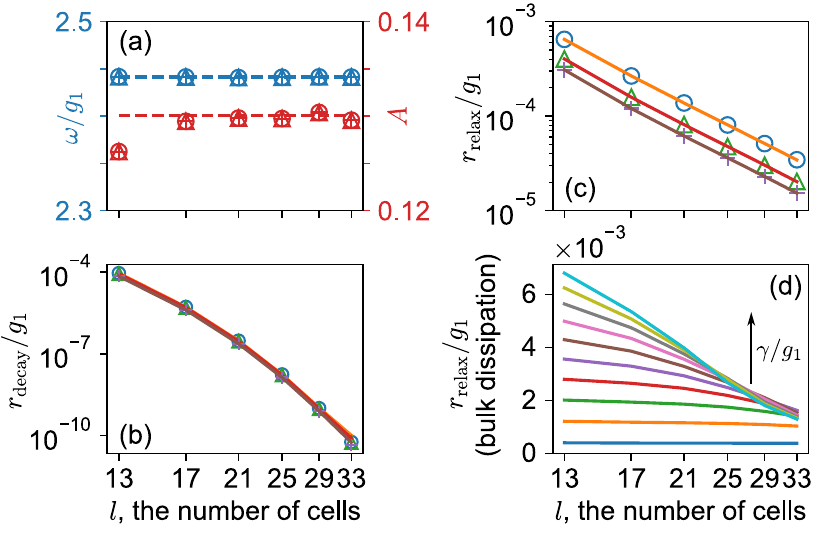}
		\caption{\quad Amplitudes $A$, frequencies $\omega$, decay rates $r_\mathrm{decay}$ and relaxation rates $r_\mathrm{relax}$ of synchronization as functions of cell numbers $l$ in the generalized four-band off-diagonal AAH model.
		(a) Synchronization frequencies $\omega$ and amplitudes $A$ for $\gamma/g_1=1$ (circles), $\gamma/g_1=2$ (triangles), and $\gamma/g_1=3$ (crosses), with $g_2/g_1=0.7$.
		Dashed lines are theoretical results in the thermodynamic limit.
		(b) Decay rates of synchronization modes fitted with $f(l)=a + bc^l$ (solid lines).
		(c) Relaxation rates for central-site dissipation fitted with $f(l)=a + b/(l+c)^d$ (solid lines).
		(d) Relaxation rates with bulk dissipation for uniformly increasing $\gamma/g_1$ from 0.002 (bottom) to 0.038 (top).
		 }\label{fig:sync_rate}
	\end{figure}

	\emph{Synchronization rate and oscillation amplitude.---}
	In small-sized systems, the synchronization between edges exhibits a notable decay over time due to the failure to meet synchronization conditions~\cite{sm}.
	The decay rate $r_\mathrm{decay}$ is proportional to the wavefunction density at the central sites with dissipation, which diminishes exponentially with the number of cells $l$ as $|g_2^2/g_1^2|^l$.
	Consequently, the synchronization has a prolonged lifetime as the number of cells increases.
	We plot in Fig.~\ref{fig:sync_rate}(a) the oscillation amplitude and frequency of the left edge site as functions of dissipation strength and the number of cells for the generalized four-band AAH model.
	For comparison, we also present the expected results in the thermodynamic limit, where the amplitude is given by
		$A = (1-g_2^2/g_1^2)^2/2$
	and independent of the dissipation strength $\gamma$~\cite{sm}.
	The consistency between finite-size results and theoretical predictions indicates that the amplitude and frequency of the synchronization are unaffected by the dissipation strength and converge to a constant as $l$ grows.
	This behavior contrasts with the findings from previous work~\cite{PhysRevLett.129.250601}, where the synchronization amplitude scales inversely with the length of the chain due to its reliance on bulk wavefunctions.
	By diagonalizing the Lindblad superoperators, we extract the decay rate of the synchronization mode, determined by the smallest modulus of the real part of eigenvalues with a nonzero imaginary part, which is also known as the spectral gap (or the asymptotic decay rate)~\cite{PhysRevA.86.012116}.
	In Fig.~\ref{fig:sync_rate}(b), we depict the decay rates over different values of $\gamma$ and $l$, fixing $g_2/g_1=0.7$, and fit the data with an exponential function $a + bc^l$.
	The fitting result $c=0.49$ aligns well with $g_2^2/g_1^2$.
	The exponential closing of the spectral gap is also observed in other systems with Anderson localization~\cite{PhysRevB.106.064203,Prosen2008}.

	Since in quantum dissipative systems with local interactions the propagation speed of the information is constrained by the Lieb-Robinson velocity~\cite{PhysRevLett.104.190401}, synchronizing two edges requires a time proportional to the system's size, given that dissipation is only applied to the central sites.
	The relaxation rate $r_\mathrm{relax}$ is set by the smallest modulus of the real part of eigenvalues excluding those associated with the synchronization modes.
	In Fig.~\ref{fig:sync_rate}(c), we plot the corresponding relaxation rates as a function of $l$ for different dissipation strengths $\gamma$.
	We observe that the relaxation rate scales as $1/l^{\alpha}$ with $\alpha \in [2,3]$ for different noise strengths, which
	is consistent with the scaling of the gap of an XY model with boundary dissipation~\cite{PhysRevLett.101.105701,PhysRevE.92.042143}.
	To boost the relaxation rate, we consider an alternative setting with the same Hamiltonian, extending dissipation to a segment of $(N+1)/2$ sites from site $(N+3)/4$ to site $(3N+1)/4$.
	The synchronization modes remain protected in the thermodynamic limit, as the distance between the edge of the chain and the boundary site of the dissipation region increases with $N$.
	However, the relaxation rate shows more complex behavior and undergoes a scaling transition observed in the XY model with the bulk dissipation~\cite{PhysRevE.92.042143}.
	As shown in Fig.~\ref{fig:sync_rate}(d), the relaxation rate remains independent of $l$ when $l$ is below a critical value $l_c$, then decreases as $1/l^\alpha$ beyond the critical point, with $\alpha=2$ from the data fitting, exhibiting a faster relaxation rate compared with Fig.~\ref{fig:sync_rate}(c) and the $r_\mathrm{relax}\propto l^{-3}$ scaling reported in the previous work~\cite{PhysRevLett.129.250601}.

\begin{figure}
		\centering
		\includegraphics[width=\linewidth]{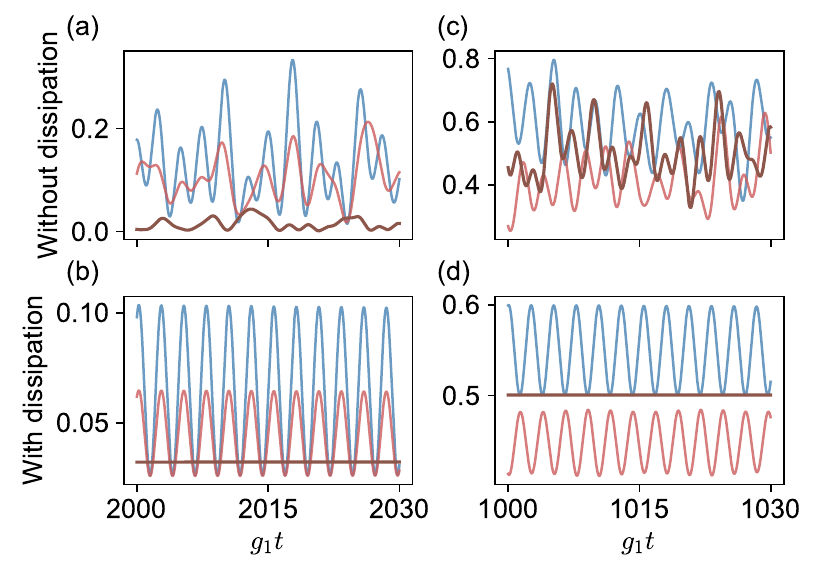}
		\caption{\quad Robustness of diagonal synchronization in the generalized four-band off-diagonal AAH model.
		The legends are the same as those in Fig.~\ref{fig:generalized}(c).
		(a)(b) Population evolution at edge and bulk sites with next-nearest-neighbor hopping $\sum_j(g_3c_j^\dagger c_{j+2} + \mathrm{h.c.})$ where $g_3/g_1=0.1$. 
		The initial state is chosen as the same as that in Fig.~\ref{fig:generalized}(c).
		(c)(d) The initial state is chosen as a random product state.}
		\label{fig:robust}
	\end{figure}

	\emph{Robustness.---}
	We now test the robustness of the diagonal synchronization in the last scenario against symmetry-broken terms.
	The synchronization grounded on the edge states persists as long as the chiral and reflection symmetries are preserved.
	Such symmetry can be broken explicitly by the next-nearest-neighbor (NNN) hopping term.
	To verify the stability of the synchronization under perturbation, we add an NNN hopping term $\sum_j(g_3c_j^\dagger c_{j+2} + \mathrm{h.c.})$ into the Hamiltonian.
	As shown in Fig.~\ref{fig:robust}(a) and (b), the evolution of the populations at the two edge sites are still synchronized under dissipation with the initial state chosen as in Fig.~\ref{fig:generalized}(c).
	We also verify that the synchronization perseveres even when disorder is introduced in the NN coupling strength within the bulk~\cite{sm}.
	This robustness originates from the resilience of the topological edge states to perturbations.

	Apart from the perturbations in the Hamiltonian, the synchronization is also robust to different choices of initial states.
	We prepare the initial state as a random product state $\otimes_{j=1}^N(\cos\theta_j + e^{i\phi_j}\sin\theta_jc_j^\dagger)|0_j\rangle$ where $\theta_j$ and $\phi_j$ are uniformly sampled from $[0, \pi)$ and $[0, 2\pi)$, respectively.
	Figure~\ref{fig:robust}(c) and (d) show that the synchronization between edge sites is established with the dissipation evolving from a random state, whereas the corresponding evolution remains uncorrelated in the absence of dissipation.

	\emph{Conclusion.---}
	We have demonstrated that synchronization between edge sites occurs in the generalized AAH models exposed to dissipation.
	In the diagonal AAH model, we have observed the synchronization of off-diagonal correlations between edge sites despite the lack of symmetries. 
	Synchronization between on-site populations is also realized in both the off-diagonal AAH model and a generalized four-band AAH model with additional chiral and reflection symmetries. 
	The synchronization amplitude and frequency converge to steady values which are independent of the dissipation strength in the thermodynamic limit.
	We also show that bulk dissipation applied to the central half of the chain can accelerate the relaxation while maintaining the synchronization mode.
	Furthermore, we reveal that the synchronization mode is robust against the symmetry-breaking terms, such as NNN interactions, and random initial states owing to the power of topology.
	Since our approaches relies solely on the spatial distribution of edge states, it can be readily extended to incorporate alternative dissipation processes such as substituting dephasing noise with particle loss~\cite{sm} and applied to any topological system that hosts edge states.
	The generalized AAH model can be implemented in optical lattices or superconducting circuits~\cite{Roati2008,Wang2024}.
	The dephasing channel can be simulated by introducing engineered noise into the lattice potentials or by modulating the fluxes in Josephson junctions within superconducting systems~\cite{Li2023,tao2024,sm}.
	Our protocol also holds practical potential in constructing long-range synchronization networks~\cite{PhysRevLett.129.063605} and communication based on synchronization~\cite{Cuomo1993,Colet1994,Argyris2005}.

	\begin{acknowledgments}
	The codes for numerical simulation are available from the corresponding author upon request.
	We acknowledge the fruitful discussions with Shang Liu and Anton Frisk Kockum.
	This research was financially supported by the Knut and Alice Wallenberg Foundation through the Wallenberg Center for Quantum Technology (WACQT).
	\end{acknowledgments}


%

\end{document}